# Magnetostructural Transformation and Magnetoresponsive Properties of $MnNiGe_{1-x}Sn_x$ Alloys

Enke Liu, Yin Du, Jinglan Chen, Wenhong Wang, Hongwei Zhang and Guangheng Wu

Beijing National Laboratory for Condensed Matter Physics, Institute of Physics, Chinese Academy of Sciences, Beijing 100190, PR China

**The martensitic and magnetic phase transformations in $MnNiGe_{1-x}Sn_x$ (0 ≤ x ≤ 0.200) alloys were investigated using X-ray diffraction (XRD), differential thermal analysis (DTA) and magnetization measurements. Results indicate that the increasing Sn substitution in $MnNiGe_{1-x}Sn_x$ results in (i) decrease of martensitic transformation temperature from 460 to 100 K and (ii) conversion of AFM spiral to antiparallel AFM strcuture in martensite. Based on these, the remarkable magnetic-field-induced PM/spiral-AFM and FM/AFM magnetostructural transformations and, large positive and negative magnetocaloric effects are obtained. The magnetoresponsive effects of $MnNiGe_{1-x}Sn_x$ alloys are enhanced by Sn substitution. A structural and magnetic phase diagram of $MnNiGe_{1-x}Sn_x$ alloys has been proposed.**

**Key words: Magnetic-entropy change, martensitic transformation, magnetoresponsive materials, MnNiGe**

## I. Introduction

Recently, the magnetic equiatomic MM'X (M, M' = transition metal, X = Si, Ge, Sn) compounds with first-order martensitic transformations has become more attractive due to their remarkable magnetoresponsive properties, including magnetic-field-induced shape memory effects [1-3] and magnetocaloric effects [4-7]. As one important system, MnNiGe crystallizes a hexagonal $Ni_2In$-type structure ($P6_3/mmc$, 194) and martensitically transforms at 470 K to the orthorhombic TiNiSi-type structure (*Pnma*, 62) [8,9]. At $T_N^M$=346 K, the martensite phase enters an antiferromagnetic (AFM) state from paramagnetic (PM) state [8]. In martensite phase, the magnetic moments with 2.3 $\mu_B$ are only localized on Mn atoms and form an AFM spiral structure [8,9]. Previous study [8] showed that, in high applied magnetic fields, the spiral structure will transit to a canted ferromagnetic (FM) one and ferromagnetically saturate at about 10 Tesla. Besides, reducing the separations of Mn atoms by substituting smaller size Si for Ge in MnNiGe [10] also enables the AFM martensite to transit to a ferromagnet. These studies all indicate instability of this AFM spiral magnetic strcuture in MnNiGe martensite. Here, this instability is expected to be manipulated to enhance the phase-transition-based magnetoresponsive properties. In this study, we substituted larger size Sn for Ge in MnNiGe to probe the phase-transition behaviors in terms of the crystallographic and magnetic structures, based on which to further investigate the magnetic-field-induced shape memory effects and magnetocaloric effects in $MnNiGe_{1-x}Sn_x$ alloys.

## II. Experimental Procedure

Polycrystalline samples of $MnNiGe_{1-x}Sn_x$ ($x$=0, 0.020, 0.035, 0.050, 0.075, 0.080, 0.085, 0.090, 0.095, 0.100, 0.200) were prepared by arc-melting raw metals in high-purity argon atmosphere. The ingots were melted four times for well alloying. The ingots were annealed in evacuated quartz tube with Ar gas at 1123 K for five days and cooled with furnace to room temperature. The room-temperature structures were identified by powder x-ray diffraction (XRD) with Cu-$K_\alpha$ radiation. No impurity phase was detected in all samples. Magnetization measurements were carried out on powder samples using a superconducting quantum interference device (SQUID). The differential thermal analysis (DTA) method with heating and cooling rate of 2.5 K/min is also used to determine the martensitic transformation data. Isothermal $M(B)$ curves for field-induced martensitic metamagnetic transformation and magnetic-entropy change ( $\Delta S_m$ ) were measured in fields up to 5 T upon cooling (heating).

## III. Results and Discussion

Low-field thermomagnetization curves, DTA and XRD analysis have been used to determine the structural and magnetic information (not shown here). According to the measured data, a structural and magnetic phase diagram has been proposed in Fig. 1. Sn substitution for Ge leads to a regular decrease of the martensitic transformation temperature ($T_t$) from 460 to about 100 K and of the transformation hysteresis from 50 to 10 K, which means that larger separations of Mn atoms in $MnNiGe_{1-x}Sn_x$ system stabilize the austenite phase. In the content range of $x$>0.095, the system shows $Ni_2In$-type austenite phase without visible martensitic transformation. Similar to the $Mn_{1-x}CoGe$ [2] and MnNiGe:Fe [3] systems, for $MnNiGe_{1-x}Sn_x$ system, the Néel temperature ($T_N^M$) of the martensite phase and the Curie temperature ($T_C^A$) of austenite both keep at 350 and 177 K, respectively, in the present Sn-substitution content. Between the $T_N^M$ and $T_C^A$, there exists a temperature window in which a martensitic transformation from PM to spiral AFM state occurs without measurable $T_N^M$ and $T_C^A$, corresponding to the Sn-substitution content of 0.025<$x$<0.085. With increasing Sn substitution, the magnetic structure of martensite transits from spiral AFM to an AFM state with more-antiparallel moment. Thus, for 0.085≤$x$≤0.095, after the spontaneous magnetization at $T_C^A$, the sample experiences a martensitic transformation from FM austenite to AFM martensite.

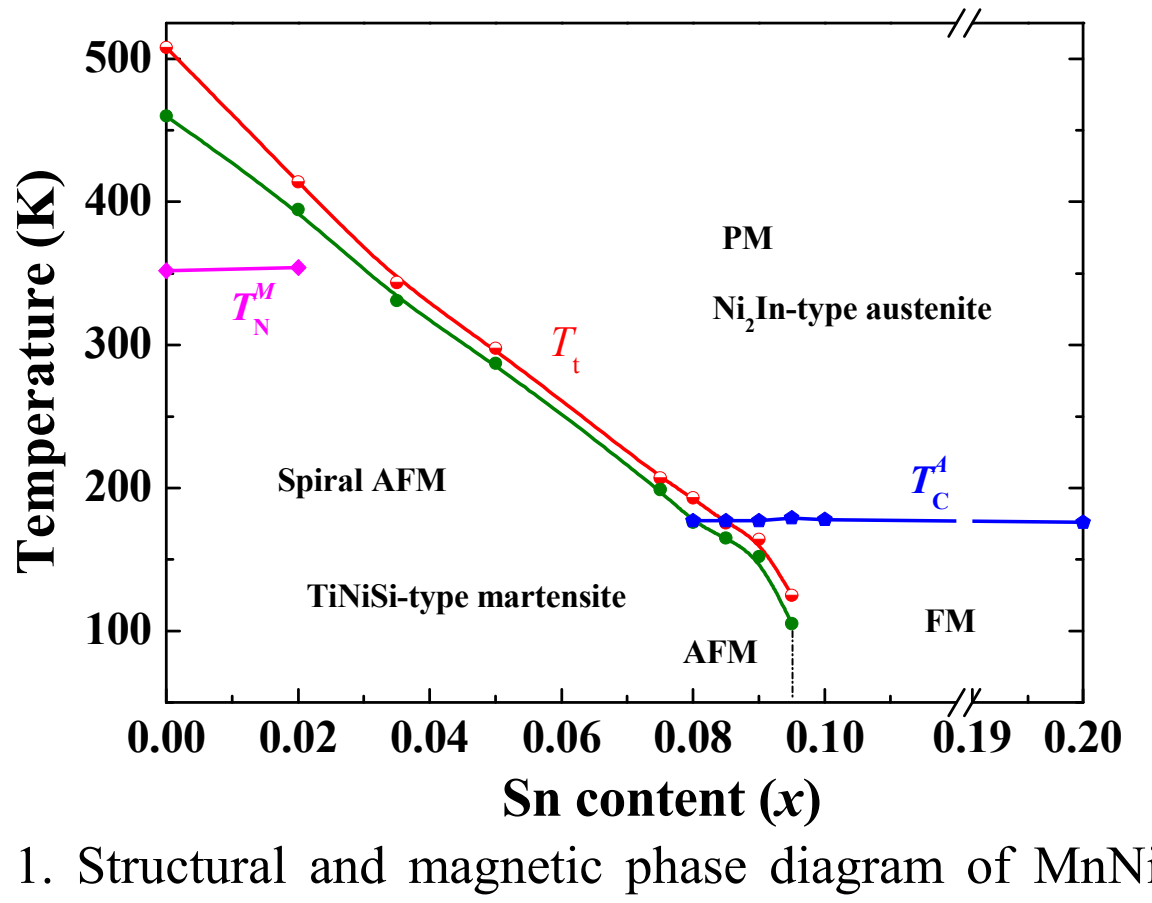

Fig. 1. Structural and magnetic phase diagram of $MnNiGe_{1-x}Sn_x$ alloys.

Figure 2a shows the thermomagnetization curves of $MnNiGe_{1-x}Sn_x$ (0.020≤$x$≤0.095) in a field of 5 T. Consistent with Fig. 1, the martensitic transformation is decreased down to the magnetic zone (below the $T_N^M$), in which the magnetic and structural transformation can coincide together. More importantly, the thermomagnetization value of produced martensite is also decreased by increasing Sn substitution.

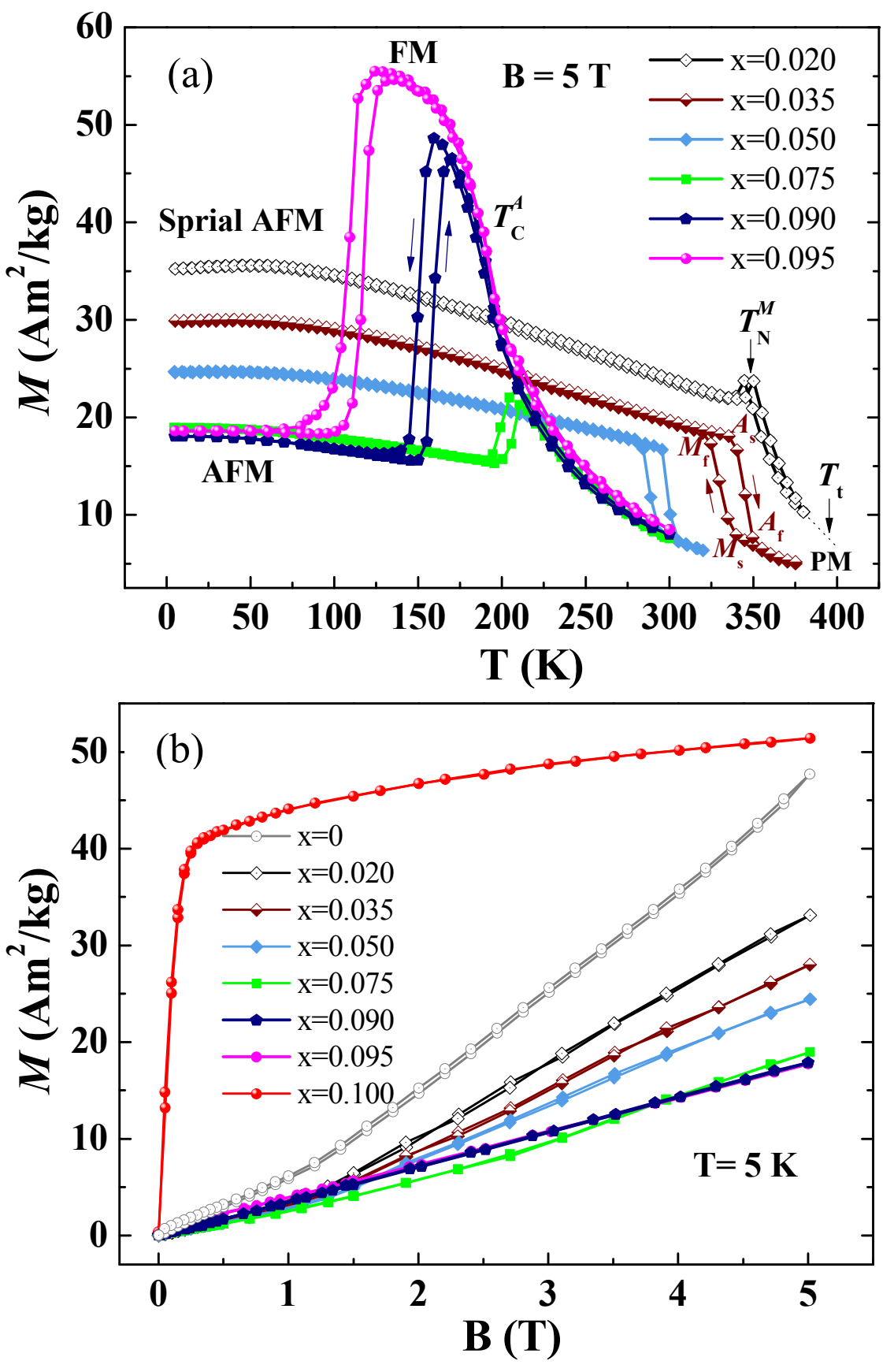

Fig. 2. Thermomagnetization curves of $MnNiGe_{1-x}Sn_x$ (0.020≤x≤0.095) in a field of 5 T (a) and magnetization curves of $MnNiGe_{1-x}Sn_x$ (0.020≤x≤0.100) at 5 K in the fields up to 5 T (b).

Within the temperature window (also see Fig. 1), one can see a PM/spiral-AFM martensitic transformation with a magnetization increment of about 10 $Am^2$/kg. For $x$≥0.075, the austenite enters into FM state due to the spontaneous magnetization at $T_C^A$ and thus shows a maximal FM magnetization of 55 $Am^2$/kg before the phase transformation, meanwhile, the magnetization of martensite has been decreased to below 19 $Am^2$/kg due to the more-antiparallel AFM state. Accordingly, a FM/AFM martensitic transformation with increasing magnetization reduction has been observed. For $x$=0.095, especially, the magnetization difference between two phases has reached to 36 $Am^2$/kg, which may benefit the magnetic-field-induced martensitic transformation.

Figure 2b shows the magnetization curves of $MnNiGe_{1-x}Sn_x$ (0.020≤$x$≤0.100) at 5 K in the fields up to 5 T. All the curves correspond to the magnetization of martensite phase, while the curve of $x$=0.100 to the austenite phase since the martensitic transformation in this composition disappears so that the curve behaves a FM magnetizing process with a magnetization of 52 $Am^2$/kg at 5 T. The Sn-free martensite shows a magnetization of about 50 $Am^2$/kg at 5 T due to the metamagnetization from spiral AFM to a canted FM state, which has a good agreement with Ref. [8]. Accordant with Fig. 2a, the magnetization of martensite is regularly decreased to 19 $Am^2$/kg with increasing Sn substitution. One can see that Sn substitution for Ge in MnNiGe results in a transition of the spiral AFM to a more-antiparallel AFM state, as also shown in Fig. 1. This explains the increasing magnetization difference between martensite and austenite of $MnNiGe_{1-x}Sn_x$ with increasing Sn substitution. As previously reported [10], in the isoelectronic system $MnNiGe_{1-x}Si_x$, created by substituting smaller size Si for Ge in MnNiGe, the AFM martensite was converted to a complete FM state. The FM exchange interactions between Mn atoms (moments) are favorable under decreased atomic separations. Inversely, an increment of Mn atomic separations will prefer AFM couplings, which is evidenced in this study by substituting larger size Sn for Ge in MnNiGe. This comparison also confirms that the MnNiGe martensite locates at a critical point on which a strong instability of magnetic structure emerges. Thus it can be easily manipulated in terms of magnetism by changing the Mn atomic separations.

Figure 3a shows magnetization isotherms of $x$=0.090 across the successive second-order and first-order transformations in the fields up to 5 T. In Fig. 3a, between 153 and 165 K, the S-shape metamagnetization curves above about 1 T indicate a remarkable magnetic-field-induced reverse martensitic transformation from AFM martensite to FM austenite. Figure 3b shows the magnetization curves across the spontaneous magnetization at $T_C^A$, corresponding to a second-order magnetic transition. The magnetic-entropy changes, based on the first- and second-order phase transformations is derived from the magnetization curves using Maxwell relation, as shown in Fig. 4. For $x$=0.050, a magnetic-entropy change of -3.5 J/kgK for $\Delta B$=5 T is obtained across the transformation. The negative value is due to the decrease of magnetic-entropy during the PM/spiral-AFM transition on cooling (see Fig. 2a). For $x$=0.075, the martensitic transformation and spontaneous magnetization are successive so that a positive magnetic-entropy change of 5 J/kgK for $\Delta B$=5 T is followed by a negative one of about -1 J/kgK for $\Delta B$=5 T in a more wide temperature range. For $x$=0.090, large magnetic-entropy changes of about 18 J/kgK for $\Delta B$=5 T across the martensitic transformation and -5 J/kgK for $\Delta B$=5 T across the spontaneous magnetization are obtained. For $x$=0.0950, a larger maxima of 20 J/kgK for $\Delta B$=5 T across the martensitic transformation is obtained as the larger magnetization value of austenite. It can be seen that the magnetic-entropy change of $MnNiGe_{1-x}Sn_x$ increases with the increasing Sn substitution.

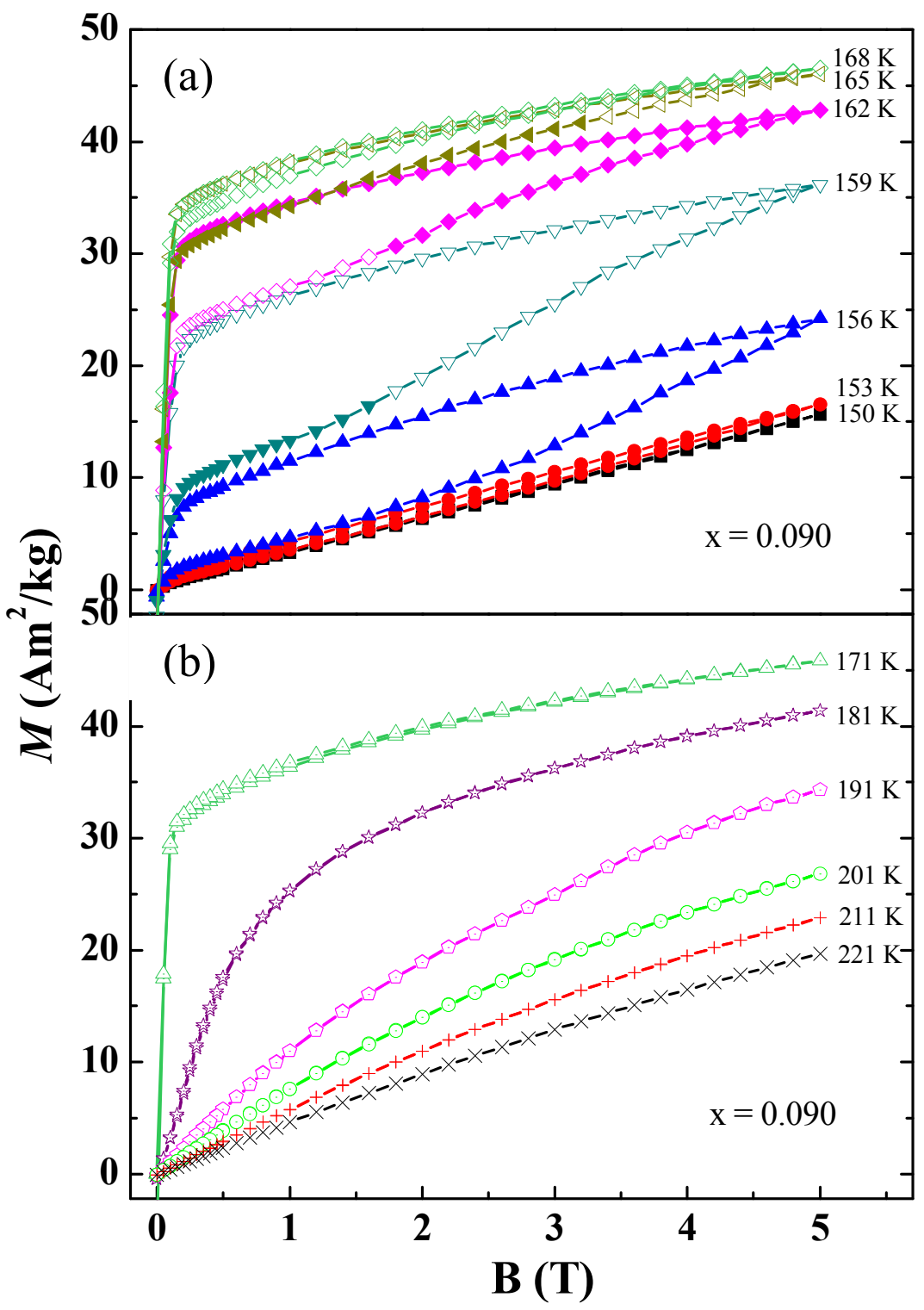

Fig. 3. Magnetization isotherms of $x$=0.090 across the successive first-order (a) and second-order (b) transformations in the fields up to 5 T.

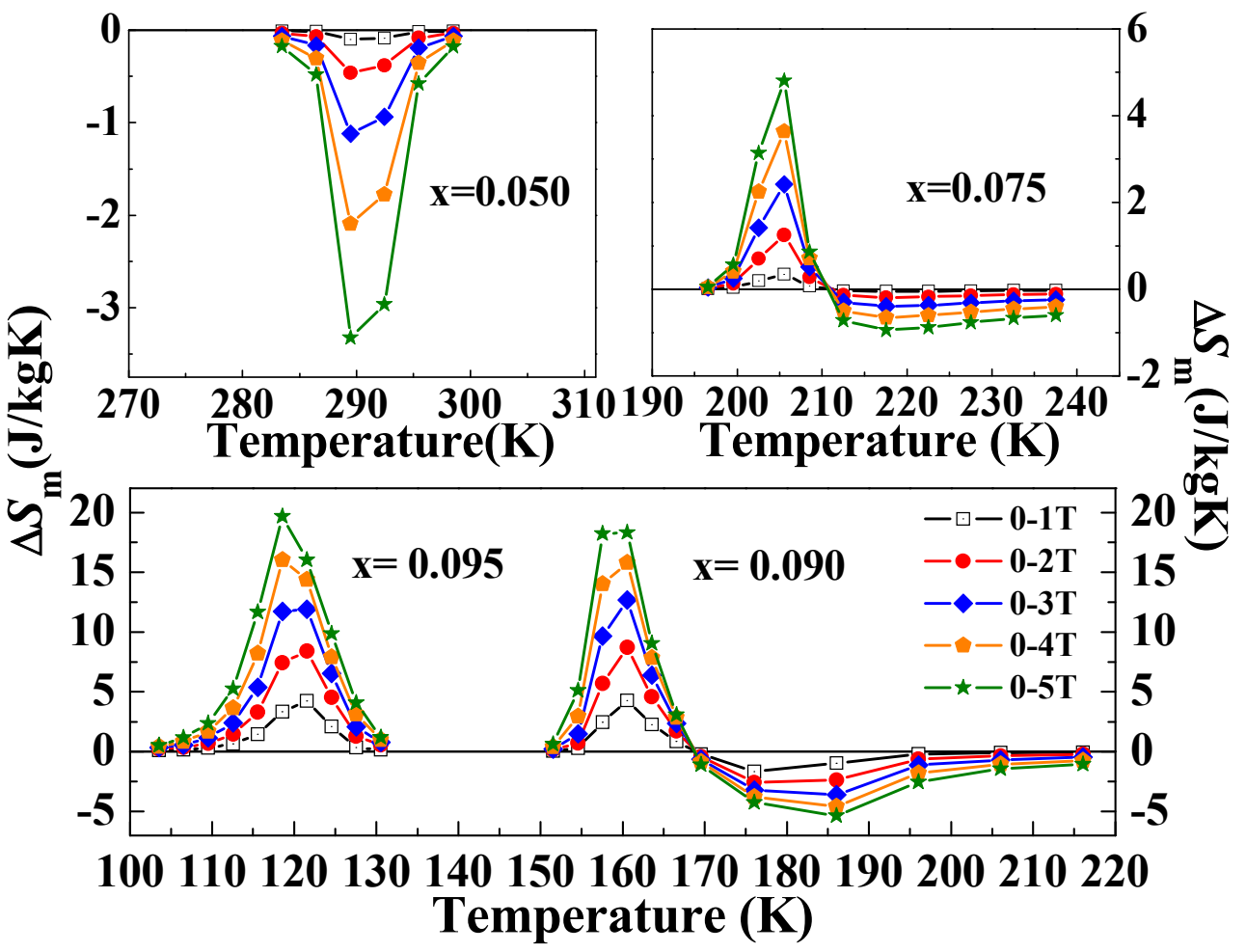

Fig. 4. Isothermal magnetic-entropy change of calculated using the magnetizing curves in different field variations of samples with $x$ = 0.050, 0.075, 0.090 and 0.095.

## IV. Conclusion

In conclusion, the martensitic transformation of $MnNiGe_{1-x}Sn_x$ alloys can be lowered from 470 to 170 K with increasing Sn substitution. The instability of spiral AFM structure of MnNiGe martensite can be manipulated to be converted to a more-antiparallel AFM state by the larger size Sn atoms. These two achievements result in a PM/spiral-AFM and FM/AFM magnetostructural transformations, based on which large magnetocaloric effects connected with the field-induced martensitic transformation are obtained. The magnetoresponsive effects of $MnNiGe_{1-x}Sn_x$ alloys are enhanced by Sn substitution. A structural and magnetic phase diagram of $MnNiGe_{1-x}Sn_x$ alloys has been proposed.

## Acknowledgment

This work was supported in part by the National Natural Science Foundation of China in Grant Nos. 51031004 and 51021061.

Corresponding author: Wenhong Wang (e-mail: wenhong.wang@iphy.ac.cn).